\documentstyle[12pt,epsf]{article}
\advance\hoffset by -1.1truecm
\advance\voffset by -1.0truecm

\def\be{\begin{equation}}
\def\ee{\end{equation}}
\def\ba{\begin{eqnarray}}
\def\ea{\end{eqnarray}}
\def\v#1{\vert #1 \rangle}

\newcommand{\R}{\mbox{I \hspace{-0.82em} R}}
\newcommand{\N}{\mbox{I \hspace{-0.82em} N}}

\newcommand{\x}{{\bf x}}
\newcommand{\p}{{\bf p}}

\newcommand{\sn}{\smallskip\newline}
\newcommand{\mn}{\medskip\newline}

\newcommand{\mbo}{{\mbox{ }}}

\begin{document}
\title{Minimal Length Uncertainty Relation and Ultraviolet
Regularisation}
\author{Achim Kempf\thanks{Research Fellow of 
Corpus Christi College in the University of Cambridge, $\mbox{
\qquad \qquad \qquad \qquad}$  
$\mbox{ }$ \quad a.kempf@amtp.cam.ac.uk}\\ \\
Department of Applied Mathematics \& Theoretical Physics\\
University of Cambridge\\ 
Cambridge CB3 9EW, U.K,\\  \\
Gianpiero Mangano\thanks{mangano@axpna1.na.infn.it,
g.mangano@amtp.cam.ac.uk}\\ \\
INFN, Sezione di Napoli, and Dipartimento di Scienze Fisiche, \\
Universit\`a di Napoli {\it Federico II}, \\ 
I-80125 Napoli, Italy} 

\date{}
\maketitle
\begin{abstract}
Studies in string theory and quantum gravity suggest the existence of 
a finite lower limit $\Delta x_0$ to the possible resolution of distances,
at the latest on the
scale of the Planck length of $10^{-35}m$. Within the framework of the
euclidean path integral we explicitly show 
ultraviolet regularisation in field theory 
through this short distance structure.
Both rotation and translation invariance can
be preserved. An example geometry is studied in detail.

\end{abstract}
\vskip-17.5truecm

\hskip10truecm
{\bf DAMTP/96-102} 

\hskip10truecm
{\bf DSF-56/96}

\hskip10truecm
{\bf hep-th/9612084}
\vskip14.5truecm

\newpage

\section{Introduction} 
As has been known for long, the combination of 
relativistic and quantum effects
implies that the conventional notion of distance breaks down 
the latest at the
Planck scale, which is about $10^{-35}m$. The basic argument is that 
the resolution of
small distances requires test particles of short
wavelength and thus of high energy. At sufficiently small scale,
i.e. close to the Planck scale,
the gravitational effect of the test particle's energy 
significantly disturbs the
space-time structure which was tried. 
Studies on gedanken experiments
therefore suggest the existence of a finite limit $\Delta x_0$
to the possible resolution of distances.
String theory, as a theory of quantum gravity, should allow a
deeper understanding of what could happen at 
such extreme scales.
Indeed, several studies in string theory yielded a certain
type of correction to the uncertainty relation
\be
\Delta x \Delta p \ge \frac{\hbar}{2} (1 + \beta  (\Delta p)^2 + ... )
,~~~\beta > 0
\label{ucr}
\ee
which, as is easily verified, implies a finite minimal uncertainty 
$\Delta x_0 = \hbar \sqrt{\beta}$. Therefore, 
$\Delta x_0 >0 $ can be viewed as a fuzzyness of space, or also as
a consequence of the
nonpointlikeness of the fundamental particles. It seems that, in string theory, 
intuitively, the input of more energy
does eventually no longer allow to improve the spatial resolution,
as this energy starts to enlarge the probed string.
References are e.g.
\cite{townsend}-\cite{maggiore}, and
recently \cite{lizzi}. For recent reviews, see e.g. \cite{garay,witten}.
\sn 
Using the usual definition of uncertainties, ($\vert\psi\rangle$
normalised)
\be 
(\Delta x)_{\vert \psi\rangle}
 = \langle \psi\vert(\x-\langle\psi\vert
 \x\vert\psi\rangle)^2\vert\psi\rangle^{1/2}
\label{ud}
\ee
the uncertainty relation Eq.\ref{ucr}
implies a small correction term to the commutation relation in
the associative Heisenberg algebra:
\be
[\x,\p]= i\hbar (1 + \beta \p^2 +...)
\label{1dcr}
\ee
For studies on the technical and conceptual implications of these 
and more general types of correction terms
see \cite{ak-jmp-ucr}-\cite{ak-prd-reg}.
We remark that those studies arose from work (e.g. \cite{ak-lmp-bf})
in the seemingly unrelated field of quantum groups, in which 
this type of commutation and uncertainty relations
had appeared independently (first in \cite{ixtapa}). A standard reference on
quantum groups is \cite{sm-book}.
\sn
For the general case of $n$-dimensions
it appears that no consensus has been reached in the
literature on which 
generalisation of Eq.\ref{1dcr}, i.e. which
particular correction terms to the uncertainty relations
could arise as a gravity effect in the ultraviolet,
or as a string effect.
Let us therefore here consider small correction
terms of a general form $(\x_i^\dagger = \x_i, \p_i^\dagger=\p_i)$
\be
[\x_i,\p_j] = i\hbar (\delta_{ij} + \beta_{ijkl} \p_k \p_l+...)
\label{one}
\ee
with the coefficients $\beta_{ijkl}$ (and also possible terms of 
higher power in the $\p_i$)
chosen such that the corresponding uncertainty relations
imply a finite minimal uncertainty $\Delta x_0 >0$. 
We will for simplicity normally assume
$[\p_i,\p_j] = 0$, but we  
allow $[\x_i,\x_j] \ne 0$. We will generally refer to
such modified commutation relations as describing 
a `noncommutative geometry'. 
Let us keep in mind however that it is the correction terms to the
$\x,\p$ commutation relations which induce $\Delta x_0 >0$. 
A noncommutativity of the $\x_i$
will not be necessary for the appearance of a finite minimal uncertainty
$\Delta x_0$.
\sn
In short, 
the key mechanism which leads to ultraviolet regularisation 
in the presence of a minimal uncertainty $\Delta x_0$ is the following:
\sn
On ordinary geometry, i.e. with the ordinary commutation relations
underlying,
the states of maximal localisation are position eigenstates 
$\vert x\rangle$, for which the uncertainty in position vanishes. 
Crucially, these maximal localisation states
are nonnormalisable. Therefore, their scalar product is not a
function but the Dirac $\delta$-distribution 
$\langle x\vert x^\prime\rangle = \delta(x-x^\prime)$. As is well known,
(for a recent reference see \cite{oleaga}),
in the formulation of local interaction in field theory it is the 
ill-definedness of the product of these and related distributions which give
rise to ultraviolet divergencies.
\sn
A finite minimal uncertainty $\Delta x_0$ will yield
normalisable maximal localisation states, and thereby
regularise the ultraviolet. 
More precisely, as we will see, there exist geometries, i.e. 
generalised commutation relations 
of the type of Eqs.\ref{one} such that there exists a minimal 
uncertainty $\Delta x_0 >0$, with the vectors of maximal localisation
$\vert x^{ml}\rangle$ obeying:
\be
\langle x^{ml} \vert x^{ml}\rangle =1, \qquad
\Delta x_{\v{x^{ml}}} = \Delta x_0, \qquad \quad \langle x^{ml}\vert
\x \vert x^{ml}\rangle = x \quad \mbox{ with } \quad x \in \R
\ee
It follows that due to their normalisability, the scalar product 
\be
\tilde{\delta} (x,y)
:= \langle x^{ml}\vert y^{ml}\rangle
\label{dt}
\ee
is a function rather than a distribution. 
\smallskip\newline
A simple example is the one-dimensional case of Eq.\ref{1dcr} with 
$\beta > 0$ and no higher order corrections. For this case
the scalar product of the maximal localisation states has
been calculated in \cite{ak-gm-rm-prd}:
\be
\tilde{\delta}(x,y) =
\frac{1}{{\pi}} \left( 
\frac{x-y}{2\hbar\sqrt{\beta}}  
 - \left(\frac{x-y}{2\hbar \sqrt{\beta}}\right)^3
 \right)^{-1}
 \sin\left(\frac{x-y}{2\hbar\sqrt{\beta}}{\pi}\right)
\label{deltabeta}
\ee 
Note that the poles of the first factor are cancelled by zeros of the
sine function, so that $\tilde{\delta}$ is a 
regular function. For a graph see Fig.3 in \cite{ak-gm-rm-prd}.
The analogous result for the case with also a finite
minimal uncertainty in momentum has been worked out in \cite{ak-hh-1}.
\sn
We consider it to be an attractive feature of this short distance structure
that it will not require the breaking of translation and rotation
invariance, while also being compatible with possible 
(e.g. quantum group-) generalisations
of these symmetries. Also, this regularisation will not require to cut
momentum space.
\sn
A general approach for the formulation of quantum field theory on 
such geometries has been developed in \cite{ft}, with a
general result on infrared regularisation in \cite{ak-prd-reg}, and
preliminary results on ultraviolet regularisation in \cite{ak-jmp-reg}.
Our aim here is to show the general mechanism, both
abstractly and explicitly, 
by which a minimal uncertainty in position 
regularises the ultraviolet, 
i.e. we show how a $\Delta x_0$ could indeed 
provide a natural ultraviolet cutoff in quantum field theory. 
\mn
While we will focus here on commutation relations which 
induce a finite $\Delta x_0 >0$, the general framework does allow for 
generic commutation relations. Let us therefore also mention some
of those studies which suggest such more general commutation relations.
\sn
For example, the approach by Doplicher et. al. \cite{doplicher}
suggests the existence of specific corrections to the $\x,\x$ commutation 
and uncertainty relations. One of the arguments there is that
the improvement of a position measurement in one direction 
ultimately requires a delocalisation 
in orthogonal directions, in order to reduce the gravitationally
disturbing energy density of the probing particle.
A possible noncommutativity of the position operators was probably first
discussed in \cite{snyder}, developing a line of thought which has 
been followed since, mainly by Russian schools, see e.g.\cite{kadys}.
In the context of noncommuting position operators, see also \cite{ahlu}.
Other studies, e.g. \cite{gac}, suggest a length dependence
of the minimal uncertainty in length measurements. 
Correction terms specifically to the $\p,\p$ commutation relations have been
discussed e.g. in \cite{townsend,banach}.
The approach of `generalised quantum dynamics' by
S. Adler \cite{adler}
allows for generic commutation relations and a possible
generalisation of the underlying Hilbert space to a quaternionic or
octonic space. In this approach the ordinary canonical commutation
relations have been derived as a first order approximation in a
statistical averaging process \cite{adler-horwitz}. 
\sn
Further, it should also 
be of interest to apply the noncommutative geometric
concepts developed in \cite{connes}, in particular to 
study the modifications to the 
differential and integral calculus over such generalised quantum
phase spaces.
\sn
We note that, technically,
the appearance of correction terms to the canonical 
commutation relations can generally also be viewed as
a nontrivial and nonunique 
change of generators from the $\x_0,\p_0$ which obey
$[\x_0,\p_0]=i\hbar$ to new
sets of generators. Examples for such algebra homomorphisms $\rho$
for the case of Eq.\ref{1dcr} are for example
$\rho : \x_0 \rightarrow \x=
\x_0 + \beta \p_0\x_0\p_0, ~
\p_0 \rightarrow \p_0$, or also $\rho : \x_0 \rightarrow 
\x =\x_0, ~ \p_0
\rightarrow \p = \beta^{-1/2}\tan(\p_0\beta^{1/2})$.
\sn
The reason why a slight change
in the commutation relations is able to introduce a drastically
new short distance structure is not only that expectation values of 
a function of operators generally do not equal the function of
the expectation values. Technically, the reason is 
of course that algebra
homomorphisms $\rho$ which change the commutation
relations of the generators
are necessarily noncanonical transformations,
i.e. unlike symmetries,
the $\rho$ cannot be implemented as unitary (nor as antilinear antiunitary)
transformations.
Unitaries $U$ generally preserve 
any chosen commutation relations, say $h(\x,\p)=0$, since
$h(\x,\p)=0\Rightarrow h(\x^\prime,\p^\prime)=
h(U\x U^\dagger,U\p U^\dagger)=U h(\x,\p) U^\dagger =0$.
Thus, any change in the commutation relations
introduces new features into the theory,
such as the appearance 
of a $\Delta x_0 > 0$, which we will here focus on.

\section{General framework}
\subsection{Partition Function}
Let us consider the example of euclidean 
charged scalar $\phi^4$-theory,
in its formulation on position space
\be
Z[J] := N \int D\phi\mbo e^{-\int d^4x \mbo [
\phi^* (-\partial_i \partial^i + m^2 c^2)\phi 
 + \frac{\lambda}{4!}(\phi \phi)^*\phi \phi - \phi^*J - J^*\phi] }
\ee
with $N$ a normalisation factor.
Fourier transformation allows to express the action functional
in momentum space, which is of course 
to choose the plane waves as a Hilbert 
basis in the space of fields which is formally being summed over. 
Equivalently, the action functional can be
expressed in any arbitrary other Hilbert basis, such as e.g. 
a Hilbert basis of Hermite functions.
In fact, it is not necessary to specify any choice of basis.
Fields can be identified as vectors in the representation space $F$ of the 
associative Heisenberg algebra ${\cal A}$ 
with the canonical commutation relations:
\be 
[\x_i,\p_j] = i\hbar \delta_{ij} \qquad i,j = 1,...,4
\label{cr}
\ee
Since the functional analytic structure is analogous to the 
situation in quantum mechanics, 
we formally extend the Dirac notation for states to fields, i.e. 
$\phi(x) = \langle x\vert \phi\rangle$
and
$\phi(p) = \langle p\vert \phi\rangle$.
We recall that, via
$\langle x\vert p\rangle = (2\pi\hbar)^{-2} \exp(ixp/\hbar)$,
the $\hbar$ which appears in the
Fourier factor $e^{ixp/\hbar}$ 
of the transformation from position to momentum space
stems from the $\hbar$ of Eq.\ref{cr}.
Of course, the simple quantum mechanical 
interpretation of fields $\vert \phi\rangle$
and in particular of
the position and momentum operators of Eq.\ref{cr} does not simply
extend, due to the relativistically necessary 
existence of anti-particles, see \cite{fdl}.
However, this formulation clarifies the functional analytic
structure of the action functional \cite{ft,ak-prd-reg}:
\be
Z[J] = N \int D\phi \mbo e^{ - \frac{l^2}{\hbar^2} \mbo \langle 
\phi \vert  
\mbo\p^2 + m^2 c^2\mbo\vert \phi \rangle \mbo
- \frac{\lambda l^4}{4!} 
\langle \phi * \phi \vert \phi * \phi \rangle \mbo
+ \langle \phi \vert J\rangle + \langle J\vert \phi\rangle }
\label{dn}
\ee
The pointwise multiplication $*$ of fields 
is crucial for the description of local interaction.
It maps two fields onto one field, i.e.
$* ~ :  F \otimes F \rightarrow F $ and it normally reads:
\be 
* =  
\int d^4x ~\vert x\rangle \otimes \langle x\vert \otimes \langle x \vert
\label{pmo}
\ee
so that, in our notation:
\begin{equation}
(\phi_1 * \phi_2)(y) = \langle y\vert \phi_1*\phi_2\rangle =
\int d^4x~\langle y\vert x\rangle~\langle x\vert\phi_1\rangle~\langle
x\vert\phi_2\rangle = \phi_1(y) \phi_2(y)
\end{equation}
On general geometries we read Eq.\ref{pmo} with the $\vert x\rangle$
denoting the vectors of maximal localisation i.e.
we are integrating over the position expectation values of the maximal 
localisation vectors:
\be 
* = 
\int d^4x~ \vert x^{ml}\rangle \otimes \langle x^{ml}\vert 
\otimes \langle x^{ml} \vert
\label{li}
\ee
In Eq.\ref{dn}, in order to make the units more transparent, 
we introduced an arbitrary 
unit length $l$, so that the fields $\vert \phi\rangle$
become unitless. $l$ could trivially also be reabsorbed in
the definition of the fields. 
As is easily seen, on ordinary geometry the vectors $\vert x\rangle$
have units $length^{-2}$, so that
$\vert\phi_1 * \phi_2\rangle$ has units $length^{-2}$, implying that
the coupling constant $\lambda$ (of the unregularised theory) is unitless. 
As is to be expected in a regularised situation,
this changes on general geometries with normalisable maximal 
localisation vectors. Due to $\langle x^{ml}\vert x^{ml}\rangle=1$
the $\vert x^{ml}\rangle $ do not carry units, so that the coupling $\lambda$
is no longer unitless.
\smallskip\newline
We recall that on ordinary geometry
the position eigenvectors are the maximal localisation vectors, implying
that the application of the definition Eq.\ref{pmo} 
for $*$ in the partition function
describes the maximally local interaction.
The apparent `nonlocality' introduced in Eq.\ref{li} is of the size
of the minimal position uncertainty in the underlying geometry.
Within the framework, physical processes, including measurement processes, 
obey the uncertainty relations.
We therefore conclude that
the so-defined interactions are observationally 
strictly local since the apparent nonlocality could not be observed -
due to the fuzzyness $\Delta x_0$ of the underlying geometry.
\smallskip\newline
In our formulation of field theories on noncommutative geometries
we will stick to the abstract 
form of the action functional and the partition function,
as e.g. given in Eq.\ref{dn}, i.e. we will not introduce
any changes "by hand" into the form of the action functional.
The switching on of corrections to the underlying geometry will automatically
manifest itself in the explicit form of the resulting Feynman rules.
The correction terms to the
commutation relations induce modifications to the
action of the operator $(\p^2 + m^2)$, and to the properties of
the maximally localised fields $\vert x^{ml}\rangle$, which 
will both crucially enter into the Feynman rules.
\sn
We remark that, as a new feature, some generalised commutation
relations will have nontrivial unitarily nonequivalent representations,
as the well-known theorem by v. Neumann no longer applies. It has been 
suggested that such cases could correspond to a noncommutative geometric
analog of geometries with horizons or nontrivial topology \cite{banach}.

\subsection{Feynman rules}
\label{sec22}
For explicitness, let us specify some arbitrary
Hilbert basis $\{\vert n\rangle\}_n$ in the space $F$ of fields
on which the generalised commutation relations are represented.
While this basis can be
continuous, discrete, or generally a mixture of both,
we here use the convenient notation for $n$ discrete. We recall that
the discreteness or continuousness of the choice of basis is
unrelated to the issue of regularisation.
$F$ is separable even in the case of the ordinary commutation relations, i.e.
discrete Hilbert 
bases (such as the Fock basis) also exist on ordinary 
geometry. We remark that on ordinary geometry, and e.g.
`on position space', the situation is slightly subtle since the
propagator and the vertex are then distributions. 
The situation will become simpler for $\Delta x_0> 0$, 
as the distributions will turn into regular functions.
\sn
Fields, operators and $*$ are expanded in 
the $\{\vert n\rangle\}$ basis as 
\be \phi_n = \langle n\vert \phi \rangle \qquad 
\mbox{and }\qquad 
(\p^2 +m^2c^2)_{nm} = \langle n \vert \p^2 + m^2c^2 \vert m\rangle
\ee
and
\be
* = \sum_{n_i} L_{n_1,n_2,n_3} \vert n_1\rangle \otimes
\langle n_2\vert \otimes \langle n_3 \vert
\ee
Thus
\be
\vert \phi * \phi^\prime\rangle = \sum_{n,m,r} L_{nmr} 
\langle m\vert\phi\rangle\langle r\vert\phi^\prime\rangle\vert n\rangle
\ee
i.e. 
\be
(\phi*\phi^\prime )_n = L_{nrs}\phi_r\phi^\prime_s
\ee
In this Hilbert basis the partition function Eq.\ref{dn}
thus reads, summing over repeated indices:
\be
Z[J] = N \int_F D \phi \mbo e^{ -\frac{l^2}{\hbar^2}\mbo
\phi_{n_1}^* (\p^2 +m^2c^2)_{n_1n_2} \phi_{n_2}
-\frac{\lambda l^4}{4!} L^*_{n_1n_2n_3}
L_{n_1n_4n_5} \phi^*_{n_2} \phi^*_{n_3} \phi_{n_4}\phi_{n_5}
+ \phi^*_n J_n + J^*_n \phi_n }
\ee
Pulling the interaction term in front of the path integral,
completing the squares, and carrying out the gaussian integrals
yields 
\be
Z[J] = N' e^{-\frac{\lambda l^4}{4!} L^*_{n_1n_2n_3} L_{n_1n_4n_5}
\frac{\partial}{\partial J_{n_2}}
\frac{\partial}{\partial J_{n_3}}
\frac{\partial}{\partial J^*_{n_4}}
\frac{\partial}{\partial J^*_{n_5}}}
\mbo
e^{-\frac{\hbar^2}{l^2} J^*_n (\p^2 +m^2c^2)^{-1}_{nm} J_m}
\label{al1}
\ee
We can therefore read off the Feynman rules for the propagator and the vertex
 
\be
G_{nm} = \left(\frac{\hbar^2/l^2}{\p^2+m^2c^2}\right)_{nm}, \qquad
\Gamma_{rstu} = - \frac{\lambda l^4}{4!} L^*_{nrs}L_{ntu}
\label{al2}
\ee
Note that the earlier arbitrarily introduced constant $l$ drops out
of the Feynman rules since each vertex attaches to four propagators.
\smallskip\newline
Explicitly, Eq.\ref{li} yields the structure constants:
\be
L_{n_1,n_2,n_3} = \int d^4x ~ \langle n_1\vert x^{ml}\rangle
\langle x^{ml}\vert n_2\rangle \langle x^{ml}\vert n_3 \rangle
\label{sc}
\ee
On ordinary geometry, we recover with $\vert x^{ml}\rangle
= \vert x\rangle$, and e.g. choosing the position representation
$\vert n\rangle = \vert x \rangle$:
\begin{equation}
L^{(\Delta x_0=0)}_{x,x^\prime,x^{\prime\prime}}
= \delta^4(x-x^\prime) \delta^4(x-x^{\prime\prime}) 
\end{equation}
In the general case with $\Delta x_0>0$, as we said, 
the coupling constant picks up
units. We can however still define a unitless $\lambda$ 
by splitting off suitable factors of $l$. 
Let us also choose $l=\Delta x_0$.
Any other choice for $l$ would amount to a redefinition of the 
coupling constant $\lambda$.
\smallskip\newline
As abstract operators, i.e.
without specifying a Hilbert basis in the space of fields, 
the free propagator and the lowest order vertex then read,
using the definition Eq.\ref{dt}:
\be
G = \frac{\hbar^2}{(\Delta x_0)^2 ({\bf p}^2+m^2c^2)}
\label{fr1}
\ee
\begin{equation}
\Gamma = - \frac{\lambda}{4!} \int 
\frac{d^4x~d^4y}{(\Delta x_0)^{8}}~~
\tilde{\delta}^4(y^{ml},x^{ml})~~
\vert y^{ml}\rangle \otimes \vert y^{ml}\rangle
\otimes
\langle x^{ml}\vert\otimes\langle x^{ml}\vert
\label{fr2}
\ee
We can now use the Feynman rules Eqs.\ref{fr1},\ref{fr2} to explicitly check
for UV regularisation on
noncommutative geometries $\cal{A}$ with $\Delta x_0>0$.
\subsection{Regularisation}
\label{2.3}
Let us first consider the tadpole graph (see Fig1).
$$ $$
\epsfysize=0.8in \centerline{\epsfbox{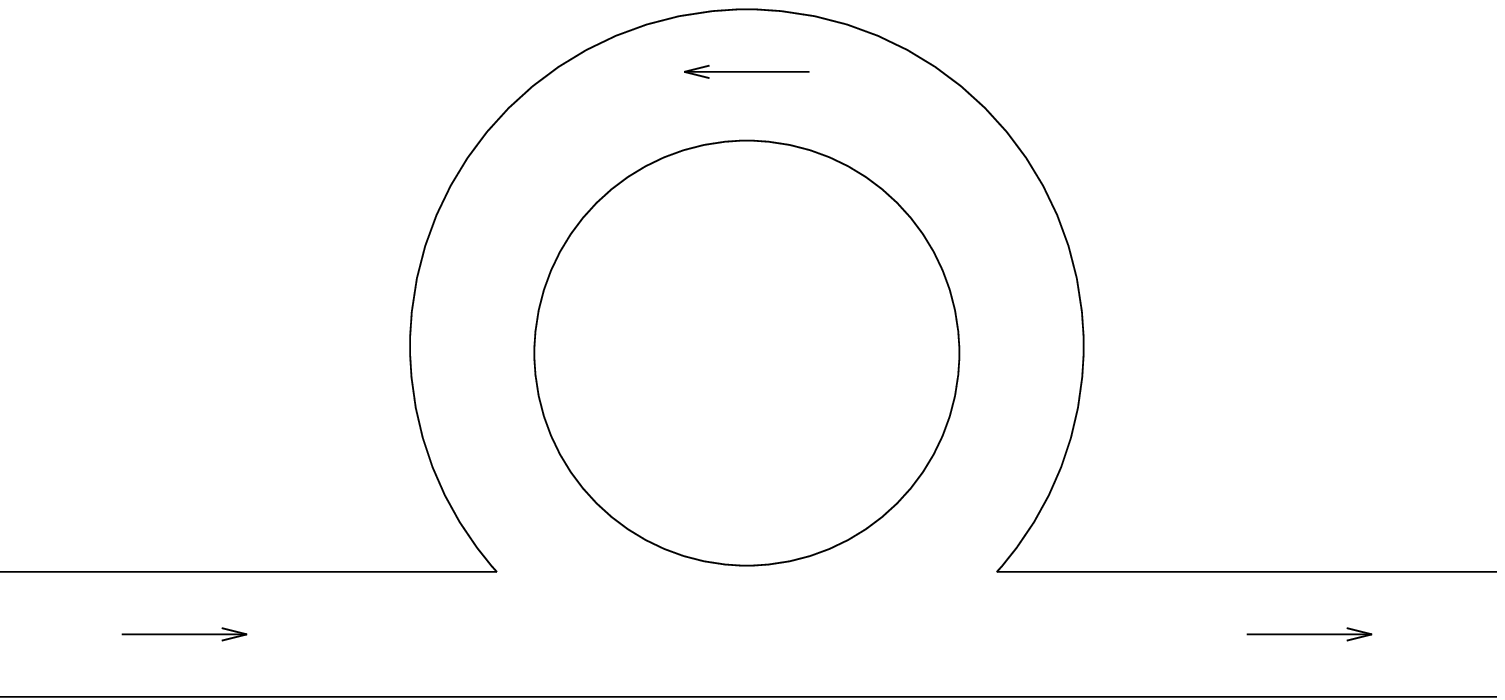}}
\vskip0.3cm
\centerline{\small \it Fig1: The tadpole graph. The notation is meant to
indicate $\Delta x_0>0$, }
\centerline{\small \it i.e. the fuzzyness of space-time, or the 
particles' nonpointlikeness.} 
$ $
\mn 
Using Eqs.\ref{al1},\ref{al2},\ref{sc}, or directly Eqs.\ref{fr1},\ref{fr2},
yields its expression as an operator:
\be
\frac{2\lambda \hbar^2}{(4!)^2 (\Delta x_0)^2}~\int\frac{d^4x~d^4y}{
(\Delta x_0)^8}~~\tilde{\delta}^4(x,y)~~
\langle x^{ml}\vert\frac{1}{\p^2+m^2c^2} \vert y^{ml}\rangle~~
\vert y^{ml}\rangle \otimes \langle x^{ml}\vert
\label{tadpole1}
\ee
As is well known, ordinarily
this graph is quadratically divergent for large momenta.
On position space the divergence, or rather the
ill-definedness of this graph, arises 
not through the large scale integrals, but instead
at short distances, i.e. as $x \rightarrow y$. 
\smallskip\newline
On our noncommutative geometries this graph is however well defined:
Due to the normalisability
of the maximal localisation vectors, their scalar 
product $\tilde{\delta}^4$ is a function bounded by 1, rather than
a distribution.
In the second factor, which consists of matrix elements of the propagator, 
the operator $(\p^2+m^2c^2)^{-1}$ is bounded. Therefore, again due to the
normalisability of the $\vert x^{ml}\rangle$ also these
matrix elements are bounded functions of $x$ and $y$. 
Thus the short-distance divergence is indeed removed on the 
noncommutative geometry.

In the case $m=0$ the operator $1/\p^2$ is unbounded, which, as is
well known, can lead to infrared divergencies at large distances.
A relevant question in this context is of course whether on a 
geometry with a finite minimal uncertainty in momentum 
this infrared problem could be avoided. Indeed, as has been
shown in \cite{ak-prd-reg}, the existence of 
a finite $\Delta p_0>0$ implies that
the operator $1/\p^2$ is as well behaved as if it contained
a mass term, i.e. it is a bounded self-adjoint operator. 
Since we are here primarily interested in the ultraviolet behaviour,
let us in the following assume the infrared to be regularised
either through $m>0$, or e.g. through 
$\Delta p_0 >0$ (examples of noncommutative 
geometries with both, finite minimal uncertainties in position $\Delta x_0$
and in momentum $\Delta p_0$ are known, see \cite{ak-jmp-ucr}).
\smallskip\newline
The tadpole graph could of course have been avoided by normal ordering the
interaction lagrangian. Let us therefore 
consider the further example of the normally 
logarithmically divergent `fish' graph (see Fig2).
$$ $$
\epsfysize=0.9in \centerline{\epsfbox{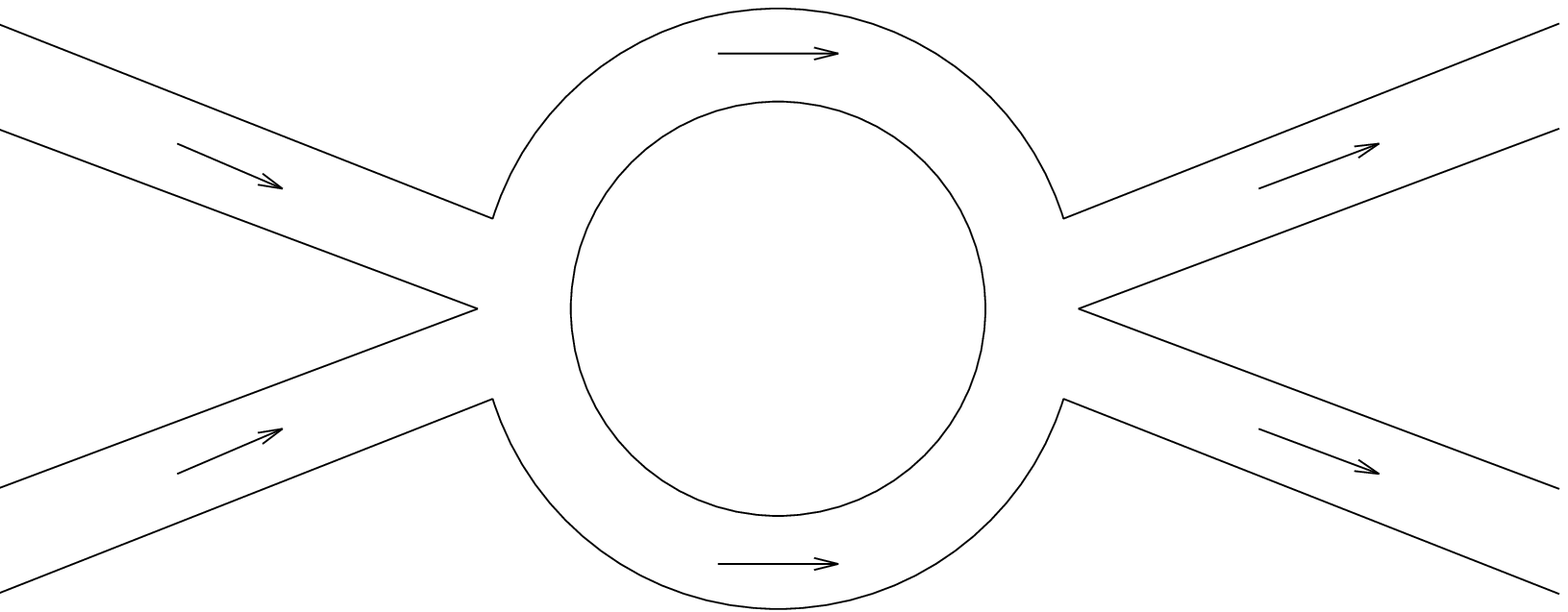}}
\vskip0.5cm
\centerline{\small \it Fig2: The fish graph. The thick 
lines for the propagator and the vertex 
are meant to}
\centerline{\it indicate the presence of a finite
minimal position uncertainty $\Delta x_0$.} 
$$ $$
It requires two vertices and two propagators:
\begin{eqnarray}
\frac{2 \lambda^2 \hbar^4}{(4!)^2 (\Delta x_0)^4} & & \!\!\!\!\!\int
\frac{d^4x_1~d^4x_2~d^4x_3~d^4x_4}{(\Delta x_0)^{16}} ~~
\langle x^{ml}_2\vert \frac{1}{\p^2+m^2c^2}\vert x^{ml}_3\rangle^2~~ \times
\nonumber \\
\nonumber \\
 & & \times~~\tilde{\delta}^4(x_1,x_2)~~ \tilde{\delta}^4(x_3,x_4)~~
\vert x_1^{ml}\rangle\otimes\vert x^{ml}_1\rangle\otimes\langle 
x^{ml}_4\vert\otimes\langle x^{ml}_4\vert
\end{eqnarray}
Ordinarily, in position space, the propagator
$\langle x_2\vert~(\p^2+m^2c^2)^{-1}~\vert x_3\rangle$ is 
divergent for $x_2\rightarrow x_3$. Nevertheless, it is well
defined as a distribution. However, its square
$\langle x_2\vert~(\p^2+
m^2c^2)^{-1}~\vert x_3\rangle^2$ is not\footnote{We remark that 
the ansatz of Differential
Renormalisation, see e.g. \cite{oleaga}, starts here by
replacing the ill defined square of the propagator
(nonuniquely) by the derivative of a well defined 
distribution, thereby introducing a length scale.}.
\sn
In contrast, since on the noncommutative geometry the
matrix elements of the propagator
$\langle x^{ml}_2\vert~(\p^2+m^2c^2)^{-1}~\vert x^{ml}_3\rangle$
are bounded, also for $x_2\rightarrow x_3$,
arbitrary high powers 
$\langle x^{ml}_2\vert~(\p^2+m^2c^2)^{-1}~\vert x^{ml}_3\rangle^r$,
$r\in \N$, 
are also well defined functions of $x_2$ and $x_3$.
Again, the short distance structure is found to be regularised.
\smallskip\newline
In fact, it is obvious that the short distance structure of 
all graphs is regularised, since in arbitrary graphs at most finite powers
of matrix elements of the propagator, and powers of $\tilde{\delta}$
can appear, which both are now bounded regular functions.
\smallskip\newline
We should note, however, that although we have seen that the 
ultraviolet divergencies 
are absent, we cannot generally exclude that a
noncommutative geometry could introduce new types of divergencies.
This will have to be investigated case by case.

\subsection{External Symmetry}
The one-dimensional uncertainty relation Eq.\ref{ucr}
has no unique $n$-dimensional generalisation. Therefore,
any particular choice for the corrections to the
commutation relations in $n$ dimensions
will require motivation from string theory or quantum gravity.
There is also the possibility
of generalised external and internal symmetry groups (e.g.
quantum groups) at the Planck
scale, see e.g. \cite{ak-jmp-ucr,sm-cqg,oswz,luk}.
We will here not attempt to develop such arguments any further. 
Let us here instead consider the constraints which can be 
posed by requiring conventional translation and rotation invariance
of the commutation relations.
\smallskip\newline
We start with a general ansatz for 
$\x,\p$ commutation relations in $n$ dimensions
\be
[\x_i,\p_j] = i\hbar \Theta_{ij}(\p)
\label{xpnd}
\ee
where we require that only the ultraviolet is affected, i.e.
$\Theta_{ij}(p)$ shall be allowed to significantly differ from 
$\delta_{ij}$ only for large momenta.
\sn
As we said, we assume $[\p_i,\p_j]=0$. 
(We remark that it has been argued that 
if the final theory of quantum theory on curved space 
does contain momentum operators, these
should be generators of a generalised definition of translation 
on curved space, in which case $[\p_i,\p_j]=0$ would express
the absence of curvature on position space \cite{banach}.) 
\sn
The remaining                            
commutation relations among the $\x_i$ are then determined
through the Jacobi identities, yielding \cite{banach}:
\be
[\x_i,\x_j] = i\hbar~\{ \x_a,\Theta_{ar}^{-1}~\Theta_{s[i}~\Theta_{j]r,s}\}
\label{xxnd}
\ee
For simplicity we adopted the geometric notation, with
$\{,\}$ and $[,]$ standing for (anti-) commutators and with
$Q,_s = \partial/\partial p_s Q$.
\sn
We observe that the $\x,\p$
commutation relations Eqs.\ref{xpnd} are 
translation invariant in the sense
that they are preserved under the transformations
\be
\x_i \rightarrow \x_i + d_i~, \qquad 
\p_i \rightarrow \p_i, \qquad d_i \in \R, ~i = 1,...n
\label{trl}
\ee 
On the other hand, for generic $\Theta$, the commutation relations
Eqs.\ref{xxnd} are not invariant under translations, 
i.e. the generators obtained through the transformations Eqs.\ref{trl} 
do not obey Eq.\ref{xxnd}. We can, however, enforce translation invariance
by requiring $\Theta$ to yield $[\x_i,\x_j] =0$. 
We read off from Eq.\ref{xxnd}   
that a sufficient and necessary condition for
this to hold is, (summing over $i$)
\be
\Theta_{ia}\partial_{p_i}\Theta_{bc} = \Theta_{ib}\partial_{p_i}
\Theta_{ac}
\label{cvrs}
\ee
which may be viewed as expressing the absence of curvature
on momentum space, by the same arguments as above.
Of course, central correction terms may still be
added on the RHS of the $\x,\x$ commutation relations, 
without spoiling translation invariance, e.g. terms
of the form suggested in \cite{doplicher}. 
\sn
The requirement of rotation invariance further imposes
\be
\Theta_{ij}(p) =  f(p^2) ~\delta_{ij} + g(p^2) ~p_i p_j
\ee
so that Eq.\ref{cvrs} takes the form \cite{ak-osc}:
\be
g = \frac{2 f f^\prime}{f-2 p^2 f^\prime}
\label{tc}
\ee
where the prime denotes $d/dp^2$.
Under these conditions translations and rotations 
do respect the commutation relations, 
i.e. they are quantum canonical transformations,
and can indeed be implemented as unitary transformations.
The translations are given by
\be
U(d):=e^{d \cdot T}
\ee
with $[T_i,\x_j]=\delta_{ij}$, and where we denoted the scalar
product $\sum_{i=1}^n d_i T_i$ by $d \cdot T$.
Since in the `naive' definition of translations in
Eqs.\ref{trl} there is no explicitly built-in
`knowledge' of the new short
distance structure, the anti-hermitean generators $T_i$ are not given by
the $-\p_i/i\hbar$ directly. Instead, they are  
\be
T_i = \frac{\p_i}{-i\hbar f(\p^2)}
\label{tau}
\ee
as is not difficult to verify.
As a consequence of the new short distance structure
the translators $T_i$ will be found to be bounded operators, technically
as we will see (Eq.\ref{fde}), because $f$ eventually
goes linearly with $p$ for large $p$.
\sn
Analogously, rotations 
\be
U(\Theta) = e^{\Theta_{ij} M_{ij}}
\ee
are generated by the operators 
\be
M_{ij} = \frac{1}{-i\hbar f(\p^2)}~({\bf{p}}_i{\bf{x}}_j -
{\bf{p}}_j {\bf{x}}_i)
\ee
which obey
\begin{equation}
[{\bf{p}}_i,{\bf{M}}_{jk}] = \delta_{ik}{\bf{p}}_j   -\delta_{ij}{\bf{p}}_k
\end{equation}
\begin{equation}
[{\bf{x}}_i,{\bf{M}}_{jk}] =  \delta_{ik}{\bf{x}}_j   -\delta_{ij}{\bf{x}}_k 
\end{equation}
\begin{equation}
[{\bf{M}}_{ij},{\bf{M}}_{kl}] = \delta_{ik}{\bf{M}}_{jl} 
- \delta_{il}{\bf{M}}_{jk} +
\delta_{jl}{\bf{M}}_{ik} - \delta_{jk}{\bf{M}}_{il}
\end{equation}
as usually.

\section{Explicit example}
In the following we will illustrate the formalism with an 
explicit example of a noncommutative geometry.

\subsection{Choice of commutation relations}
If we require our example geometry to obey 
translation and rotation invariance, 
there still appears to be considerable freedom in choosing
the functions $f$ and, through Eq.\ref{tc}, the function $g$.
Many choices may not lead to geometries with a minimal uncertainty
$\Delta x_0 >0$. In particular, Eq.\ref{tc} indicates that
$g$ can develop singularities. A detailed
investigation into the various possibilities
is in progress \cite{ak-gm-3}. Here, in order to
obtain a well behaved example geometry we simply force there
not to appear a singularity by imposing, as the simplest choice
$(\beta > 0)$:
\be
g = \beta
\label{gde}
\ee
Thus, Eq.\ref{tc} then reads
\be
f^\prime = \frac{\beta f}{2 (f+ \beta p^2)}
\ee
which is solved by
\be
f = \frac{\beta p^2}{\sqrt{1+2 \beta p^2}-1}
\label{fde}
\ee
The Taylor expansion around the origin is well behaved:
\be
f = 1+ \frac{\beta}{2} p^2 + {\cal O}((\beta p^2)^2)
\ee
so that, if we choose $\beta$ e.g. at 
around the Planck scale $\beta^{-1/2} \approx p_{pl}$
then $f$ significantly deviates from the identity only for large
momenta of that scale.
\sn
We therefore obtain the commutation relations:
\begin{eqnarray}
[\x_i,\p_j] & = & i\hbar 
~\left(\frac{\beta \p^2}{(1+2 \beta \p^2)^{1/2}-1}~\delta_{ij}
 +\beta \p_i\p_j
\right)
\label{eg1} \\
 {[} \x_i , \x_j {]}  & = &   0  \label{eg2} \\
\quad [\p_i,\p_j] & = & 0
\label{eg3}
\end{eqnarray}
We remark that, assuming translation and rotation invariance,
the correction terms to the commutation relations are in fact unique
to first order in $\beta$:~
Eq.\ref{tc} yields $f=1+\beta/2~ p^2 
+ {\cal O}(\beta^2)$ and $g= \beta +
{\cal O}(\beta^2)$, so that
\be
[\x_i,\p_j] = i\hbar ~ \left( (1+\beta/2~ \p^2) ~ 
\delta_{ij} +  \beta \p_i \p_j + {\cal O}(\beta^2)\right)
\ee
and $[\x_i,\x_j] = 0 + {\cal O}(\beta^2), ~ [\p_i,\p_j]=0$,
which  of course coincides with what we obtain from Eqs.\ref{eg1} 
to first order in $\beta$.
\sn
We remark that concerning 
the possible choices of commutation relations it
should generally be interesting to investigate
the interplay of the technical constraints 
with the input and physical intuition from string theory and 
quantum gravity. In particular,
as follows from the relation between the
translators and the momenta, Eq.\ref{tau}, the rule for the addition
of extremely large momenta is modified through 
$(p \hat{+} k)_i = p_i f^{-1}(p^2) + k_i f^{-1}(k^2)$. There should
exist an interpretation in terms of the effects of 
gravity at the Planck scale, similar to the well known effect
of momentum nonconservation through gravity on large scales 
$(T^{\mu\nu}{}_{;\nu}=0$ rather than $T^{\mu\nu}{}_{,\nu} =0$).
This may e.g. be related to the old idea of possible curvature
in momentum space, in which a generalised
parallelogram rule for the addition of momenta has been discussed,
see \cite{snyder}, and more recently \cite{kadys}.
It has of course long been suggested that, more drastically, 
both rotation and translation invariance may be generalised
or broken at the Planck scale. Any physical intuition for this could
and should then also provide guidance for the generalisation of Eq.\ref{li}
to account for the then position (and possibly orientation) dependence
of the short distance structure of the geometry. This will at first
require a case by case study.

\subsection{Hilbert space representations}
The commutation relations 
Eqs.\ref{eg1},\ref{eg2},\ref{eg3} still find
a Hilbert space representation in the spectral
representation of the momenta $\p_i$ 
(since momentum space is still commutative and
there is no finite minimal uncertainty in momentum, $\Delta p_0=0$)
\begin{eqnarray}
\x_i.\psi(p) & = & i\hbar~\left( \left( f^\prime + p^2 g^\prime 
+\frac{n+1}{2} g\right) p_i + f\partial_{p_i} +g~ p_i p_j\partial_{p_j}
\right) \psi(p)
\\
\p_i.\psi(p) & = & p_i~ \psi(p)\\
\langle \psi_1 \vert \psi_2 \rangle & = &
\int d^np~ \psi_1^*(p) \psi_2(p) 
\end{eqnarray}
where $\psi(p)=\langle p\vert \psi\rangle$ and $\langle p\vert p^\prime
\rangle = \delta(p-p^\prime)$. $\x_i$ and $\p_i$ are symmetric operators
on the dense domain $D:= S_{\infty}$.
This representation 
holds for any choice of $f$ and $g$, as can be checked directly.
The case of commutation relations with 
general $\Theta$ is covered in \cite{banach}.
\sn
A further representation of the commutation relations 
Eqs.\ref{eg1},\ref{eg2},\ref{eg3}, 
which will prove convenient for
practical calculations, is obtained by using that
the translators $T_i$ are anti-hermitean
and have a spectral representation on the Hilbert basis 
$\{\vert {\rho}\rangle~ \vert {\rho} \in
I_n \}$ of vectors obeying 
$T_i.\vert {\rho}\rangle ={\rho}_i/i\hbar \vert {\rho} \rangle$ with
\be
I_n= \{{\rho} \in \R^n\vert~ {\rho}^2 < 2/\beta\}
\ee
i.e. the $T_i$ are bounded operators. 
The unitary transformation which maps from momentum space
to the spectral representation of the $T_i$ has the
matrix elements:
\be
\langle {\rho}\vert p\rangle = 
(1-\beta {\rho}^2/2)^{-\frac{n+1}{2}}(1+\beta {\rho}^2/2)^{1/2}~
\delta^n \left( p_i - \frac{{\rho}_i}{1-\beta {\rho}^2/2} \right)
\label{pip}
\ee
The operator representations and the scalar product
then read in ${\rho}$-space:
\begin{eqnarray}
\x_i.\psi({\rho}) & = & i\hbar~ \partial_{{\rho}_i} \psi({\rho}) \\
\p_i.\psi({\rho}) & = & \frac{{\rho}_i}{1-\beta {\rho}^2/2}~ \psi({\rho})\\
\langle \psi_1 \vert \psi_2 \rangle & = &
\int_{I_n} d^n{\rho}~ \psi_1^*({\rho}) \psi_2({\rho}) 
\end{eqnarray}
where $\psi({\rho}) = \langle {\rho}\vert \psi\rangle$ and $\langle {\rho}\vert
{\rho}^\prime\rangle = \delta^n({\rho}_i-{\rho}_i^\prime)$.
Note that, as is easy to see in this representation, 
the momentum operators $\p_i$ are still unbounded.
\sn
We still have to prove that the geometry defined through the
commutation relations Eqs.\ref{eg1},\ref{eg2},\ref{eg3}
does in fact imply a finite minimal uncertainty $\Delta x_0>0$,
rather than e.g. a discretisation of position space.
Before we do this in the next section, let us note an
important representation theoretic consequence of the
existence of a minimal uncertainty $\Delta x_0>0$:
\sn
A general argument shows that commutation relations which
imply a finite minimal uncertainty in position cannot
find a Hilbert space representation on a spectral representation
of the the position operators:
The uncertainty relations hold in all 
$*$-representations of the commutation relations.
On the other hand, as is easily
seen, e.g. in the example of Eq.\ref{ud},
an eigenvector to an observable
necessarily has vanishing uncertainty in this observable.
Thus, if the 
uncertainty relations imply a finite uncertainty in positions, they
exclude the existence of any position eigenvectors in any
physical domain, i.e. on any domain
 on which the commutation relations are represented.
In particular, in cases where $\Delta x_0>0$ and $\Delta p_0>0$
both position and momentum representations are ruled out and
one has to resort to other Hilbert bases, as e.g. in \cite{ft,ak-jmp-reg}.
\sn
To be precise, let us assume that the commutation relations 
are represented on some dense domain $D \subset H$
in a Hilbert space $H$. Ordinarily, there would exist sequences
$\{\vert \psi_n\rangle \in D\}$ with position uncertainties
decreasing to zero (e.g. Gaussian approximations to the 
position eigenvectors). In the presence of a finite $\Delta x_0>0$,
however, there exists a minimal uncertainty 
`gap', i.e. there are no vectors $\vert \psi\rangle \in D$
which would have an
uncertainty in positions in the interval $[0,\Delta x_0[$, 
so that now
\be
\exists\!\!\!/~~~ \{\vert \psi_n\rangle \in D\}: ~ \lim_{n\rightarrow \infty}
(\Delta x_0)_{\vert \psi_n\rangle} = 0
\ee
Technically, the position operators
are merely symmetric on representations $D$
of the commutation relations.
Their deficiency indices are
nonvanishing and equal, implying the existence of a family
of self-adjoint extensions in $H$, though, crucially of course,
not in $D$. This functional analytic structure was first
found in \cite{ak-jmp-ucr}.
\sn
As is easily seen, 
there do exist formal position eigenvectors in $H$:
\be
\psi_{\xi} ({\rho}) = \left[ \left( \frac{\beta }{2 \pi}
 \right)^{ \frac{n}{2} }
\frac{ n~ \Gamma ( n/2 )}{2 } \right]^{ 
\frac{1}{2} } e^{- i \xi \cdot {\rho} /
\hbar}
\label{eigenx}
\ee
Concerning 
the normalisation, recall that the surface of the $(n-1)$-dimensional
unit sphere reads $S_n = \int d\Omega_n = 2 \pi^{n/2}/\Gamma(n/2)$.
The scalar product can be calculated to be
\begin{eqnarray}
\langle \psi_{\xi} \vert \psi_{\eta} \rangle & = & 
\left[ \left( \frac{\beta }{2 \pi}
 \right)^{ \frac{n}{2} }
\frac{ n~ \Gamma ( n/2 )}{2 } \right]
\int_{I_n} d^n\rho~e^{-i(\eta-\xi)\cdot \rho/\hbar} \nonumber \\
 & = &
\left( \frac{ \sqrt{2} \hbar \sqrt{\beta} }{\vert \xi - \eta \vert}
\right)^{\frac{n}{2}} \Gamma \left( \frac{n}{2} + 1 \right)
J_{ \frac{n}{2} } \left( \frac{\sqrt{2} \vert \xi - \eta \vert}
{\hbar \sqrt{\beta} } \right)
\label{prodeigen}
\end{eqnarray}
where $J_{n/2}$ is the Bessel function of the first kind of order $n/2$.
The zeros of the scalar product determine the self-adjoint
extensions of the on $D$ densely defined $\x_i$ (for any chosen $\xi$,
all $\eta$'s such that $|\xi - \eta|$ is a zero of $J_{n/2}$
correspond to the eigenvectors of one self-adjoint extension).
However, as is readily verified, none of these vectors is 
in the domain of the $\p_i$. Thus, as is to be expected when
$\Delta x_0>0$, none of the family of self-adjoint
extensions of the $\x_i$ is in the domain of the representation
of the commutation relations. In the one-dimensional
case $n=1$ we recover the results obtained in \cite{ak-gm-rm-prd},
in particular the scalar product of the 
`formal position eigenvectors' (technically of eigenvectors
of the adjoints 
$\x_i^*$, which are not self-adjoint, nor symmetric):
\be
\langle \psi_{\xi} \vert \psi_{\eta} \rangle  = 
\frac{\hbar \sqrt{\beta}~ \sin \left( \frac{\sqrt{2} \vert \xi - \eta \vert}
{\hbar \sqrt{\beta} } \right) }{ \sqrt{2} \vert \xi - \eta \vert}  
\ee
There is, however, a
natural generalisation of the position space
representation. To this end we define a Hilbert space 
representation of the commutation relations on `quasi-position 
space', see \cite{ak-gm-rm-prd,ak-hh-1}:
\be 
\psi(x):= \langle x^{ml}\vert \psi\rangle
\ee
These quasi-position functions $\psi(x)$ are obtained 
by projecting the fields $\vert \psi\rangle$
onto the fields of maximal localisation 
$\vert x^{ml} \rangle$ and they do of course turn into the ordinary
position space representation for $\Delta x_0\rightarrow 0$.

\subsection{Maximally localised fields}
Let us now prove that $\Delta x_0>0$ by explicitly calculating the 
maximally localised fields.
\sn
As is well known the $\Delta x_i \Delta p_i$ 
uncertainty relations are derived from the positivity of the norm:
\be
\vert\vert (\x_i -\langle \x_i\rangle) + i k (\p_i-\langle\p_i\rangle)
\vert\psi\rangle\vert\vert \ge 0
\ee
Thus, the vectors on the boundary of the region allowed by the uncertainty
relations obey the squeezed state equation:
\be
(\x_i -\langle \x_i\rangle) + i k (\p_i-\langle\p_i\rangle)
\vert\psi\rangle = 0
\label{sse}
\ee
Due to the symmetry of the underlying geometry we 
do not loose generality by calculating
the maximally localised field
$\vert x^{ml}\rangle$, around the origin, i.e. with 
$\langle 0^{ml}\vert \x_i \vert 0^{ml} \rangle =0$
and $\langle 0^{ml}\vert \p_i \vert 0^{ml} \rangle =0$.
In ${\rho}$-space Eq.\ref{sse} reads
\be
\left(i\hbar \partial_{{\rho}_i} + i k \frac{{\rho}_i}{1-\beta {\rho}^2/2}\right)
\psi_k({\rho}) = 0
\label{pr}
\ee
Due to rotational symmetry, $\vert 0^{ml} \rangle$ 
can only depend on $p^2$, so that Eq.\ref{pr} becomes
\be
\partial_{{\rho}^2} \psi_k({\rho}^2) = -\frac{k}{2 \hbar
 (1-\beta {\rho}^2/2)}~
\psi_k({\rho}^2)
\ee
whose normalised solutions read $(k>0)$:
\be
\psi_k({\rho}^2) = \left[ \left( \frac{\beta}{2 {\pi}} \right)^{\frac{n}{2}}
\frac{ \Gamma \left( \frac{2 k}{\hbar \beta} + \frac{n}{2} + 1 \right)}
{\Gamma \left( \frac{2 k}{\hbar \beta} + 1 \right) } \right]^{\frac{1}{2}}
\left( 1 - \beta {\rho}^2/2 \right)^{\frac{k}{\hbar \beta}}
\ee
We can now calculate the squared uncertainty 
in position as a function of $k$:
\be
(\Delta x)^2_{\vert\psi_k\rangle} = \frac{\hbar k}{4} ~
\frac{4k +n\hbar \beta}{2k - \hbar \beta}
\ee
The minimum is reached for
\be
{k_0} = \frac{\hbar \beta}{2} \left( 
1 + \sqrt{ 1 + \frac{n}{2} }
\right)
\label{kbar}
\ee
We therefore find the finite minimal uncertainty $\Delta x_0$:
\be 
(\Delta x_0)^2 = (\Delta x)^2_{\vert\psi_{{k_0}}\rangle} =
\frac{ \hbar^2 \beta}{4} 
\sqrt{1+\frac{n}{2}} ~ \left( \sqrt{1+\frac{n}{2}}+1\right)^2
\label{sei}
\ee
The field $\vert 0^{ml}\rangle = \vert \psi_{k_0}\rangle$
of maximal localisation around the origin therefore reads, in
the ${\rho}$-representation:
\be
\langle {\rho}\vert 0^{ml}\rangle = \psi_{{k_0}}({\rho}^2) =
N^{1/2}(n)~ \left(\beta/2\pi\right)^{n/4}~
\left( 1 - \beta {\rho}^2/2 \right)^{
1/2 + \sqrt{1/4 +n/8}}
\label{mlr1}
\ee
where we defined:
\begin{equation}
N(n) := \frac{ \Gamma \left( 2+ n/2 + \sqrt{1+n/2}\right)}
{\Gamma \left(2 + \sqrt{1+\frac{n}{2}} \right)}
\label{sev}
\end{equation}
The ${\rho}$-space representation of 
the fields of maximal localisation around arbitrary
position expectation values $\xi$
now follow by translation:
\be
\langle {\rho}\vert x^{ml}\rangle = \langle {\rho} 
\vert~e^{x \cdot T}~
\vert 0^{ml}\rangle = \langle {\rho}\vert 0^{ml}\rangle
 ~e^{- i x \cdot {\rho} / \hbar}
\label{mlr2}
\ee 
Using Eq. \ref{pip}, we eventually obtain
the maximally localised fields in momentum space,
\ba
\langle p\vert x^{ml} \rangle & = &
 \left( \beta/2\pi \right)^{n/4}~
N^{1/2}(n)~
 \left( \frac{ \beta p^2 } { 1 + 2 \beta p^2
- \sqrt{1 + 2 \beta p^2} } \right)^{1/2}  \label{pxmle}\\
  \nonumber \\
 & \times & \left( \frac{ \sqrt{ 1 + 2 \beta p^2 } -1 } { \beta p^2}
\right)^{1+n/2+\sqrt{1/4+n/8}}
~ \exp \left( - i \frac{ x \cdot p}{\hbar} 
\frac{ \sqrt{1 + 2 \beta p^2} - 1} 
{\beta p^2} \right)  \nonumber 
\ea
This expression is of course also the quasi-position representation
of the plane wave with momentum $p$. 
We observe that in quasi-position space the fields can now not have
arbitrarily fine ripples. Indeed, from the argument
in the exponent in Eq.\ref{pxmle} we read off that 
for increasing momentum the wavelength in quasi-position space
only tends towards a finite minimal wavelength
\be
\lambda_0 = \pi \hbar \sqrt{2\beta}
\label{fmw}
\ee
which is reached as the momentum $p_i$ tends to infinity.
The situation is perhaps comparable
to the speed of light as a fundamental limit, in which case also
the kinematics lets the energy diverge as the
fundamental limit is approached. In the Appendix we give the
unitary transformation from the $\rho$-representation to the
quasi-position representation and back, and we
prove the completeness of the set of 
maximally localised fields. 
\subsection{Feynman rules}
As we saw in Sec.\ref{sec22}, the Feynman rules are
composed of two basic functions, related to the vertex and to
the propagator respectively: 
\be
\tilde{\delta}(x^{ml},y^{ml}) := \langle x^{ml}\vert y^{ml}\rangle
~~~~\mbox{and}~~~~ G(x,y) := \frac{\hbar^2}{(\Delta x_0)^2}
\langle x^{ml}\vert (\p^2+m^2 c^2)^{-1}\vert y^{ml}\rangle
\ee
The calculation  of $\tilde{\delta}$, i.e. of the
scalar product of maximally localised fields
is straightforward, in particular in the ${\rho}$-representation.
The result is of course independent of 
the choice of Hilbert basis. Choosing spherical coordinates:
\begin{eqnarray}
\tilde{\delta}(x-y) & = &
\langle x^{ml} \vert y^{ml}\rangle \\ \nonumber \\
 & = &  \left( \beta/2\pi \right)^{n/4}~
N^{1/2}(n)~ \int_{I_n} d^n\rho~
\left( 1 - \beta {\rho}^2/2 
\right)^{1+\sqrt{1+n/2}} e^{i (x-y) \cdot \rho/\hbar}
 \nonumber \\
 \nonumber \\ & = & 
\left( \frac{ \hbar \sqrt{2\beta} }
{\vert x - y \vert} \right)^{ 1 + \frac{n}{2} 
+ \sqrt{1+ \frac{n}{2}} }
\Gamma \left(  2 + \frac{n}{2} + \sqrt{ 1 + \frac{n}{2} } \right)
J_{1 + \frac{n}{2} + \sqrt{ 1 + \frac{n}{2} }} \left(
\frac{\sqrt{2} \vert x - y \vert}{\hbar \sqrt{\beta} } \right)
\nonumber \label{prodxieta}
\end{eqnarray} 
$$ $$
\vskip-3.7cm
\epsfysize=5.2in \centerline{\epsfbox{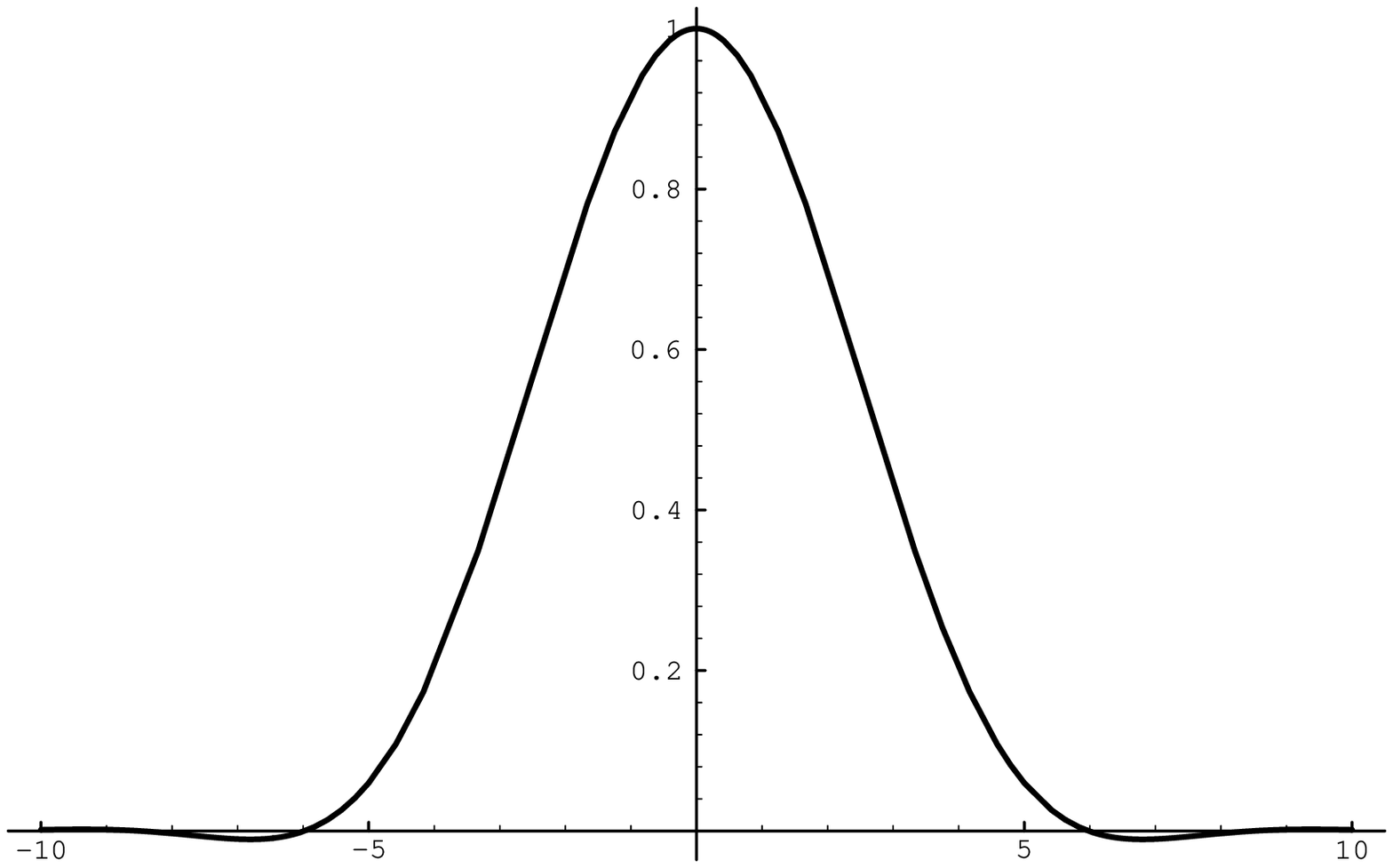}}
%
\vskip-3cm
\centerline{{\small \it Fig.3: Plot
of  the scalar product
of maximally localised fields $\tilde{\delta}(x-y)$
versus $|x-y|/ \hbar \sqrt{\beta}$.}}
\centerline{{\small \it  $\tilde{\delta}$ generalises the Dirac 
$\delta$-distribution for $\Delta x_0 >0$.}}
\vskip1.2cm
We note that $\tilde{\delta}(x-y)$ is also
the quasi-position wave function of the around $y$ maximally localised field
$\vert y^{ml}\rangle$. 
Because of the finite norm of the maximally localised fields,
$\tilde{\delta}(x-y)$ had of course to come out as a regular function.
Its graph is plotted in Fig.3. 
\sn
Recall that on ordinary geometry, i.e. when $\Delta x_0= 0$, 
the propagator in position space $G(x-y)$ can only be
defined as a distribution, and that it is the ill-definedness of
its square (as well as of higher powers) which 
gives rise to ultraviolet divergencies, see e.g. \cite{oleaga}. 
We know from Sec.\ref{2.3} that $G(x-y)$ must now 
be a well-defined function without singularities. 
Explicitly, let us consider the free propagator matrix elements:
\begin{eqnarray}
 G(x-y) & = & \frac{\hbar^2}{(\Delta x_0)^2}
\langle x^{ml} \vert (p^2 + m^2 c^2)^{-1} \vert y^{ml} \rangle  \nonumber \\
&  = &  \frac{N(n) \hbar^2}{(\Delta x_0)^2} \left( \frac{\beta}{2 \pi}
\right)^{\frac{n}{2}}
\int_{I_n} d^n \rho~
\frac{(1- \beta \rho^2/2)^{3+ \sqrt{1 + n/2}} }
{\rho^2 + m^2 c^2 (1- \beta \rho^2/2)^2}~~ e^{i(x-y) \cdot \rho/ \hbar}
\end{eqnarray}
The massive propagator cannot be simply expressed in terms of elementary or
special functions. However, for arbitrary nonvanishing mass, $G(x-y)$ can be
uniformly bounded by
\begin{equation}
\vert G(x-y)\vert \leq  \frac{ \hbar^2}{m^2 c^2 (\Delta x_0)^2}
\end{equation}
which shows that the propagator is well behaved for all distances,
in particular also for $\vert x-y\vert \rightarrow 0$.
\sn
Since the small distance behaviour is independent of the mass,
let us also consider the simpler massless propagator. Using 
spherical coordinates and introducing
the dimensionless variables
\begin
{equation}
t= \rho \sqrt{\beta/2},~~~~s=\cos \theta
\end{equation}
we obtain
\begin{eqnarray}
& & G(x-y) = \nonumber \\  
& = & \frac{N(n) \hbar^2 \beta}
{2(\Delta x_0)^2} \left( \frac{1}{ \pi}
\right)^{\frac{n}{2}} S_{n-1}
\int_0^1 dt \int_0^{\pi} d \theta ~(sin \theta)^{n-2}
t^{n-3} (1- t^2)^{3+ \sqrt{1 + n/2}}
e^{i d t cos\theta}
\nonumber \\
& = & \frac{N(n) \hbar^2 \beta }{2 (\Delta x_0)^2} \left( \frac{1}{ \pi}
\right)^{\frac{n}{2}} S_{n-1}
\int_0^1 d t \int_{-1}^1 ds ~(1-s^2)^{(n-3)/2} t^{n-3} \nonumber \\
& & ~~~~~~~~~~~~~~\times~~ (1- t^2)^{3+ \sqrt{1 + n/2}} e^{i d t s}
\end{eqnarray}
where $d := \sqrt{2}|x-y|/ (\hbar \sqrt{\beta})$.
Performing the integration over $s$ and then over $t$, 
and after some simplification, one finally obtains for $n>2$:
\begin{eqnarray}
G(x-y)  & = &
2^n ~ \frac{\left(3+\sqrt{1+n/2}\right)\left(2+\sqrt{1+n/2}\right)}
{(n-2) \sqrt{1+n/2} \left(1+ \sqrt{1+n/2}\right)^2
\left(2+n/2+\sqrt{1+n/2}\right)}  \nonumber \\
&  & \times~{}_1F_2 \left(-1 + n/2;~3+n/2+\sqrt{1+n/2},~n/2;~
-|x-y|^2/(2 \hbar^2 \beta) \right) \nonumber \\ 
\end{eqnarray}
where we have used the explicit expressions for
$(\Delta x_0)^2$ and $N(n)$ given in Eqs.\ref{sei},\ref{sev}.
\sn
In the particular case of four euclidean dimensions the last
expression can be cast in a much simpler form. For $n=4$, 
Eqs.\ref{sei},\ref{sev} for
$(\Delta x_0)^2, N(n)$ and the definition of
$\tilde{\delta}(x-y)$ yield
\begin{equation}
G(x-y)= \frac{36 + 20 \sqrt{3}}{6 + 4 \sqrt{3}}
\frac{\hbar^2 \beta}{|x - y|^2} \left(1 - \tilde{\delta}(x-y) \right)  
\end{equation}
Therefore,
the propagator can be expressed as the product of the usual zero
mass propagator and a smooth {\it cut-off} function,
which has the following behaviour for large and short distances
\begin{eqnarray}
\hbar^2 \beta \left( 1- \tilde{\delta}(x-y) \right) & \sim &
\frac{1} {8 + 2 \sqrt{3}} |x-y|^2
\left( 1 + {\cal O} \left( \left( \frac{\sqrt{2}|x-y|}
{\hbar \sqrt{\beta}} \right)^2
\right) \right), \nonumber \\
& & \mbox{for}
~~~|x-y| << \hbar \sqrt{\beta} \nonumber \\ 
 & & \\
\hbar^2 \beta \left( 1- \tilde{\delta}(x-y) \right) & \sim & \hbar^2 \beta
\left( 1 + {\cal O}
\left( \left( \frac{\sqrt{2}|x-y|}{\hbar \sqrt{\beta} }
\right)^{-7/2 -\sqrt{3}} \right) \right), \nonumber \\
& & \mbox{for}~~~ 
|x-y| >> \hbar \sqrt{\beta}\nonumber
\end{eqnarray}
In particular $G(x-y)$ is a well behaved function in the short distance
regime and tends to a finite limit for $|x-y|=0$, while for distances larger
than $\hbar \sqrt{\beta}$ it rapidly approaches the well known $|x-y|^{-2}$
behaviour of the free massless propagator on ordinary geometry.

\section{Summary and Outlook}
Studies in string theory and quantum gravity provide theoretical 
evidence for various types of correction terms
to the canonical commutation relations. Measurable effects could in 
principle appear 
anywhere between the presently resolvable scale of $10^{-18}m$
and the Planck scale of $10^{-35}m$, although expected 
close to the Planck scale.
\sn
Our aim here was to show that
the existence of even an at present unmeasurably small
$\Delta x_0$, for example at about the Planck length, could
have a drastic effect in field theory, namely by 
rendering the theory ultraviolet finite. We note that
this new short distance behaviour would truly be a 
quantum structure, in the sense that
it has no classical analog. Further, the presence of
a $\Delta x_0>0$ is
compatible with both, generalised symmetries as well as with
conventional rotation and translation symmetry.
\sn
The existence of this short distance structure would raise
a number of conceptual issues, in particular, it would lead 
to a generalised notion of local interaction. Strictly speaking,
the maximally local interaction term on generalised geometries 
is neither local nor nonlocal in the conventional sense. Instead, it 
is `observationally' local in the sense that it is local as far as
distances can be resolved on the given geometry. 
\sn
Similarly, questions such as whether the unitarity 
of time evolution is broken or conserved, or whether 
local gauge invariance is broken or conserved also 
seem not to be applicable in the usual
sense. Instead, on generalised geometries
the notions of `time evolution' or `local' gauging 
may need to be redefined, analogously to how  
local interaction is generalised into `observationally'
local interaction. This is under investigation.
\sn
There exist a number of immediate 
technical issues which need to be addressed, for example
the significance of Eq.\ref{tc} and its pole structure,
and Wick rotation. It should of course also be worth exploring
the possible usefulness of
the approach as a mere regularisation method.

\section*{Appendix}
We prove the completeness of the set of maximally localised fields
and we give the unitary transformation which connects the
$\rho$-representation with the quasi-position representation for the example
geometry considered in Section 3.

In order to see that the set of maximally localised fields 
$\{ \vert x^{ml}\rangle\}$ is complete, we use that
a set of vectors $\vert\psi_\lambda\rangle$
in a Hilbert space is complete iff from 
$\langle\phi\vert \psi_\lambda\rangle=0$,
$\forall~ \vert\psi_\lambda\rangle$ follows that $\langle \phi\vert=0$.
Consider now $\langle\phi \vert x^{ml} \rangle$:
\begin{equation} 
\langle \phi | x^{ml} \rangle =
N^{1/2}(n) ( \beta /2 \pi)^{n/4}
\int_{I_n} d^n \rho~ (1- \beta \rho^2/2)^{1/2+ \sqrt{1/4 + n/8}}
e^{i x \cdot \rho/\hbar} \phi(\rho)
\end{equation} Using the mean value theorem
\begin{eqnarray}
\langle \phi | x^{ml} \rangle &=& N^{1/2}(n) ( \beta /2 \pi)^{n/4}
(1- \beta \overline{\rho}^2/2)^{1/2+ \sqrt{1/4 + n/8}} \int_{I_n}
d^n \rho~
e^{i x \cdot \rho /\hbar} \phi(\rho)  \nonumber \\
&=& N^{1/2}(n) \frac{2}{n \Gamma(n/2)}
(1- \beta \overline{\rho}^2/2)^{1/2+ \sqrt{1/4 + n/8}} ~
\langle \phi | \psi_x \rangle  \label{mean}
\end{eqnarray}
where the $\vert\psi_x\rangle$ are eigenvectors to the 
(nonhermitean) operators $\x_i^*$ (see Eq.\ref{eigenx}) and 
$\overline{\rho}$
is some value in the open interval $]0,\sqrt{2/\beta}[$.
The set of vectors $\vert\psi_x\rangle$ is 
the collection of all eigenbases to the self-adjoint 
extensions of the $\x_i$, and is therefore overcomplete
(for the details of the functional analysis see 
\cite{ak-jmp-ucr,ak-gm-rm-prd,ak-hh-1,ixtapa}).
Further, the factor in front of the scalar product
in Eq.\ref{mean} never vanishes 
nor diverges for any $\overline{\rho}$. Thus, if
the right hand side of Eq.\ref{mean} vanishes for all $x$ this
implies $\vert\phi\rangle=0$, which had to be shown.
\sn
The completeness of the set of maximally localised fields means that
we obtain the full information on any $\vert\psi\rangle$ when collecting
its projections $\langle x^{ml}\vert \psi\rangle$ 
on the $\vert x^{ml}\rangle$, i.e. the quasi-position representation
truly represents the fields. Indeed, the mapping e.g. from $\rho$-space
to quasi-position space is invertible:
\sn
Using the explicit expressions Eq.\ref{mlr1},\ref{mlr2}
for the maximally localised fields
in the $\rho$-representation we obtain for arbitrary
$\vert \psi\rangle$ the quasi-position wave 
function $\psi(x)$ expressed in terms of its
$\rho$-representation $\psi(\rho)$ as
\begin{equation}
\psi(x) = \langle x^{ml}\vert\psi\rangle =
N^{1/2}(n) ( \beta /2 \pi)^{n/4}
\int_{I_n} d^n \rho ~(1- \beta 
\rho^2/2)^{1/2+ \sqrt{1/4 + n/8}} e^{i x \cdot \rho
/\hbar} \psi(\rho)
\label{U}
\end{equation} 
with the inverse:
\begin{equation}
\psi(\rho) =
N^{-1/2}(n) ( \beta /2 \pi)^{-n/4} (2 \pi \hbar)^{-n}
 (1- \beta \rho^2/2)^{-1/2- 
 \sqrt{1/4 + n/8}} \int d^n x ~e^{- i x \cdot \rho
/\hbar} \psi(x)
\label{U-1}
\end{equation}
Let us denote the mapping from $\rho$-space to 
quasi-position space (Eq.\ref{U}) by $U$.
The identity
$U^{-1} U = 1$ is easily verified by inserting
Eq.\ref{U} into Eq.\ref{U-1}. 
\medskip\newline
In the quasi-position representation the scalar product and the  
action of the position and momentum operators then read: 
\begin{eqnarray}
\langle \psi | \phi \rangle & = &
N^{-1}(n) ( \beta /2 \pi)^{-n/2} (2 \pi \hbar)^{-2 n}      
\int_{I_n} d^n \rho~ (1- \beta \rho^2/2)^{-1- \sqrt{1 
+ n/2}} \nonumber \\
& & \int \int d^n x ~ d^n y ~e^{i (x-y) \cdot \rho/\hbar} \psi^*(x) \phi(y)
\label{qpsp}\\
{\bf p}^i . \psi(x) & = &  - i \hbar 
\sum_{r=0}^{\infty} (\hbar^2\beta\Delta/2)^r
\frac{\partial}{\partial x_i} \psi(x) 
\label{2ndqp} \\
{\bf x}^i . \psi(x) & = &\left(x^i + i \frac{\hbar \beta}{2}
\left(1 + \sqrt{1 + \frac{n}{2}} \right) {\bf p}^i \right) \psi(x)
\label{3rdqp}
\end{eqnarray}
where $\Delta = \sum_i \partial^2/\partial x_i^2$. Note that 
the action of $\p_i$ given in Eq.\ref{2ndqp} (and also used in
Eq.\ref{3rdqp}) is well defined on quasi-position wave functions
$\psi(x)$, since they Fourier decompose into wavelengths not 
smaller than the finite minimal wavelength $\lambda_0$ (Eq.\ref{fmw}). 
In this context see also \cite{ak-gm-rm-prd} where the concept 
of quasi-position representation was first introduced.


\begin{thebibliography}{99}
\bibitem{townsend} P.K. Townsend, Phys. Rev. {\bf D15}, 2795 (1976)
\bibitem{grossmende} D.J. Gross, 
P.F. Mende, Nucl. Phys. {\bf B303}, 407 (1988)
\bibitem{amati} D. Amati, M. Ciafaloni, G. Veneziano, Phys. Lett. 
{\bf B 216}, 41 (1989)
\bibitem{konishi} K. Konishi, G. Paffuti, P. Provero, Phys. Lett. 
{\bf B234}, 276 (1990) 
\bibitem{jaeckel} M.-J. Jaeckel, 
S. Reynaud, Phys. Lett. {\bf A185}, 143 (1994)
\bibitem{maggiore} M. Maggiore, Phys. Lett. {\bf B319}, 83 (1993)
\bibitem{lizzi} F. Lizzi, N.E. Mavromatos,
Preprint OUTP-96-66-P, DSF 53/96, hep-th/9611040
\bibitem{garay} L.J. Garay, Int. J. Mod. Phys. {\bf A10}, 145 (1995)
\bibitem{witten} E. Witten, Phys. Today {\bf 49} (4), 24 (1996)
\bibitem{ak-jmp-ucr} A. Kempf, J. Math. Phys. {\bf 35} (9), 4483 (1994)
\bibitem{ft} A. Kempf, Preprint DAMTP/94-33, 
hep-th/9405067, and Czech. J. Phys. (Proc. Suppl.), {\bf 44}, 1041 (1994)
\bibitem{ak-gm-rm-prd} A. Kempf, G. Mangano, 
R.B. Mann, Phys. Rev. {\bf D52}, 1108 (1995), hep-th/9412167 
\bibitem{ak-jmp-reg} A. Kempf, Preprint hep-th/9602085, to appear in
J. Math. Phys. (1997)
\bibitem{ak-hh-1} A. Kempf, H. Hinrichsen, 
J. Math. Phys. {\bf 37}, 2121 (1996), hep-th/9510144
\bibitem{ak-prd-reg} A. Kempf, Phys. Rev. {\bf D54}, 5174 (1996),
hep-th/9602119
\bibitem{ak-lmp-bf} A. Kempf, Lett. Math. Phys. {\bf 26}, 1 (1992)           \bibitem{ixtapa} A. Kempf, Proc. XXII DGM Conf. Sept.93 Ixtapa
(Mexico), Adv. Appl. Cliff. Alg (Proc. Suppl.) {\bf (S1)} (1994)
\bibitem{sm-book} S. Majid, 
{\it Foundations of Quantum Group Theory}, CUP (1996)
\bibitem{oleaga} S.A. Pernice, G. Oleaga, Preprint hep-th/9609139
\bibitem{doplicher} S. Doplicher, K. Fredenhagen, J.E. Roberts,
Comm.Math.Phys. {\bf 172}, 187 (1995)
\bibitem{snyder} H. S. Snyder, Phys. Rev. {\bf 71}, 38 (1947)
\bibitem{kadys} V.G. Kadyshevskii, D.V. Fursaev, 
Theor. Math. Phys. {\bf 83}, 474 (1990)
\bibitem{ahlu} D.V. Ahluwalia, Phys. Lett. {\bf B339}, 301 (1994)
\bibitem{gac} G. Amelino-Camelia, J. Ellis, N.E. Mavromatos,
D.V. Nanopoulos, Preprint CERN-TH/96-143, quant-ph/9605044
\bibitem{banach} A. Kempf, Preprint DAMTP/96-10, hep-th/9603115,
to appear in {\it Quantum Groups and Quantum Spaces},
    Inst. of Math., Polish Acad. Sci., Eds: R. Budzynski, 
    W. Pusz and S. Zakrzewski, 
Banach Center Publications {\bf 40} (1997)
\bibitem{adler} S.L. Adler, Nucl. Phys.{\bf B415}, 195 (1994)
\bibitem{adler-horwitz} S.L. Adler, L.P. Horwitz, Preprint
IASSNS-HEP-96/99, TAUP 2379-96, hep-th/9610153
\bibitem{connes} A. Connes, \it Noncommutative geometry, \rm AP (1994)
\bibitem{fdl} R.P. Feynman, 
\it Dirac Memorial Lecture, `The reason for antiparticles' \rm CUP (1987)
\bibitem{sm-cqg} S. Majid, Class. Quantum Grav. {\bf 5}, 1587 (1988)
\bibitem{oswz} O. Ogievetskii, W. B. Schmidke, J. Wess, B. Zumino,
Commun. Mat. Phys. {\bf 150}, 495 (1992)
\bibitem{luk} J. Lukierski, Preprint hep-th/9610230
\bibitem{ak-osc} A. Kempf, Preprint DAMTP/96-39, hep-th/9604045
\bibitem{ak-gm-3} A. Kempf, G. Mangano, in preparation
\bibitem{gsw} M.B. Green, J.H. Schwarz, E. Witten, \it Superstring Theory \rm,
CUP, (1987)
\end{thebibliography}
\end{document}